\documentclass[twocolumn,prd]{revtex4}

\usepackage{amsfonts} \usepackage{amsmath} \usepackage{amssymb}
\usepackage{url}
\usepackage{epsf} 

\newcommand{\pa}{\partial} 

\newcommand{\scri}{\mbox{\usefont{U}{rsfs}{m}{n} I}}
\newcommand{\cM}{\mathcal{M}}

\allowdisplaybreaks

\begin{document}
\title{Numerical evolution of axisymmetric, isolated systems in
  General Relativity} 
\date{\today} 

\author{J\"org Frauendiener}
\author{Matthias Hein} 

\affiliation{Institut f\"ur Astronomie und 
  Astrophysik, Universit\"at T\"ubingen, Auf der Morgenstelle 10,
  D-72076 T\"ubingen, Germany}

\begin{abstract}
  We describe in this article a new code for evolving axisymmetric
  isolated systems in general relativity. Such systems are described
  by asymptotically flat space-times which have the property that they
  admit a conformal extension. We are working directly in the extended
  `conformal' manifold and solve numerically Friedrich's conformal
  field equations, which state that Einstein's equations hold in the
  physical space-time. Because of the compactness of the conformal
  space-time the entire space-time can be calculated on a finite
  numerical grid.  We describe in detail the numerical scheme,
  especially the treatment of the axisymmetry and the boundary.
\end{abstract}

\maketitle
\section{Introduction}

Currently, there are three main approaches to solve Einstein's
field equations using numerical methods. They differ in the way they
set up systems of equations and the appropriate initial boundary value
problems which they try to solve. The main approach is based on
variants of the ADM equations~\cite{arnowittdeser62} which are written as a
hyperbolic system of evolution equations. The IBVP is set up on
space-like hypersurfaces with boundary `far out' in a region where it
is assumed that the space-time is such that in a certain approximation
it is close to the Minkowski or Schwarzschild space-time or a
perturbation thereof. Radiation is extracted by matching to such
approximate space-times.

The second approach is based on the characteristic initial value
problem which is based on outgoing null hypersurfaces. The Einstein
equations can be written in this context in a hierarchical form which
lends itself very easily to numerical treatment. The use of outgoing
null hypersurfaces has the advantage that radiation of any type and,
in particular, gravitational radiation can be followed all the way out
to an asymptotic observer (LIGO) located at null-infinity. By using a
compactified radial coordinate it is possible to read of the radiation
at finite points.

There are some attempts to match these two approaches in the so called
Cauchy-Characteristic Matching procedure (CCM) where the two schemes
exchange appropriate information across an interface to provide
boundary conditions at a time-like hypersurface (see~\cite{winicour98:_charc_evol_match}
for a recent review).

The third approach, termed the conformal approach, is somewhat
different. While the previous methods set up equations directly for
the physical quantities in the physical space-time in this approach
one formulates the so called `conformal field equations' for
quantities on a conformally related space-time, which state in an
indirect way, that the physical Einstein tensor vanishes. This
indirection introduces some freedom which is used to compactify the
space-time in so that it is possible to hold the entire
(semi-global) problem on a finite grid. In this way one can solve a
global problem with finite resources. Since the whole space-time is
accessible one can also read off radition without further
approximation. Due to these properties the conformal approach is well
suited to study global properties of space-times.

All these approaches have certain more or less severe
draw-backs. Common to all of them is the fact that the constraints
which can be satisfied initially very accurately are violated during
the course of the evolution and this `constraint violation' seems to
grow in all codes with an exponential rate.

The standard Cauchy approach suffers from the fact that the outer
boundary is very difficult to handle. This is due to a lack of
understanding of the appropriate IBVP in the sense that there is no
boundary condition known, which would be physically meaningful
(modelling the fact that the space-time should be asymptotically flat)
as well as numerically stable. The characteristic codes crash when the
underlying outgoing null hypersurfaces form caustics and self
intersections. In the conformal approach there are no (known)
difficulties out of principle. The problems arising there as we will
point out in this work (see also a recent discussion
in~\cite{frauendiener02:_some}) are of a numerical nature or due to the complexity
of the equations.

The conformal approach has been pioneered by H\"ubner who implemented
the first code treating a spherically symmetric space-time with a
scalar field coupled to
gravity~\cite{huebner94:_method_calc_sing_spti}. He also succeeded to
evolve general initial data which were close enough to flat data to
evolve into a global space-time with a regular
$i^+$~\cite{huebner01:_from_now} (see also~\cite{husa:_probl_succes}).

We present in this work a code to solve the conformal field equations
for space-times which are axisymmetric with a regular axis. In
contrast to previous
work~\cite{frauendiener98:_numer_hivp_i,frauendiener98:_numer_hivp_ii,frauendiener99:_init_data_ps}
where the unphysical assumption was made that there are no fixed
points now we assume the existence of a regular axis which implies
that we can now treat \emph{physically relevant} isolated systems.

Our motivation for developing a 2D instead of a full blown 3D code is
that we want to have a code with reasonable turn around times that
allows us to study the effects of changing various properties such as
gauges, boundary conditions etc. within a reasonable time. Futhermore,
in contrast to a 3D code a 2D code allows us to compute the
space-times with a much higher resolution.

The plan of this paper is as follows. In sec.~\ref{sec:setting} we
review the basics of the conformal approach, write down the
conformal field equations and discuss the implementattion of the
axisymmetry. Sec.~\ref{sec:numerics} is devoted to the description of
the numerical methods used in the code. In particular, we discuss the
evolution algorithm, the boundary conditions and the issues arising
from the axisymmetry. In sec.~\ref{sec:results} we present the results
obtained so far and conclude with a summary in
sec.~\ref{sec:summary}.

\section{The General Setting}
\label{sec:setting}

\subsection{The Conformal Field Equations}

In this section we first give a short introduction into the conformal
picture and provide the conformal field equations. Then we introduce
the hyperboloidal initial value problem for the conformal field
equations by making a $3+1$-decomposition with respect to a frame into
a system of constraint equations and a system of symmetric hyperbolic
evolution equations.

We then describe how to implement the axisymmetry with special
emphasis on the regularity of the axis. The symmetry reduces the
problem from a 3D to a 2D problem, but does not lead to a significant
reduction of components. We therefore assume additionally, that the
Killing vector field $\xi^{a}$, which generates the axisymmetry is
orthogonal to the hypersurfaces of constant $\phi$, where $\phi$ is
the parameter along the orbits of $\xi^{a}$ (i.e.,
$\xi(\phi)=1$). This gives us an additional discrete symmetry, which
reduces the number of components from 65 for the general equations to
35 independent components.

Suppose we have a vacuum solution $(\widetilde{\cM}, \tilde{g})$ of
the Einstein equations, which we will call the `physical
space-time'. This space-time is asymptotically flat at null-infinity
(`light-like asymptotic flatness') and hence describes an isolated
general relativistic system, if it admits the following construction.

There exists a smooth Lorentz manifold $(\cM,g_{ab})$ and a
diffeomorphism $\Psi: \widetilde{\cM}\longrightarrow \cM$, such that:
\begin{enumerate}
\item $\Psi(\widetilde{\cM}) \subset \cM$ and $\Psi(\widetilde{\cM})$ has a
  smooth boundary, $\pa{\Psi(\widetilde{\cM})}=\scri$ with
  $\scri = \scri^{+} \cup \scri^{-}$ and $\scri^{+} \cap
  \scri^{-} = \emptyset$.
\item there exists a smooth function $\Omega$ on $\cM$, such that:
  $g_{ab}=\Omega^{2}(\Psi^{-1})^{*}\tilde{g}_{ab}$ and $\Omega \neq 0$
  on $\Psi(\widetilde{\cM})$,
\item on $\scri$ the conformal factor $\Omega=0$ and $d\Omega \neq 0$,
\item \label{geod_bed} every null geodesic in $\widetilde{\cM}$ has a
future endpoint on $\scri^{+}$ and a past endpoint on $\scri^{-}$,
\item the Ricci tensor $\tilde{R}_{ab}$ of the physical metric
  $\tilde{g}_{ab}$ vanishes in a neighborhood of $\scri$, $\tilde{R}_{ab}=0$.
\end{enumerate}
This construction states, that the physical manifold
$(\widetilde{\cM}, \tilde{g})$ is conformal to the interior of the
`unphysical' manifold $(\cM,g)$. In order to simplify the notation, we
will discard further on the diffeomorphism $\Psi$ and identify the
points $p \in \widetilde{\cM}$ and $\Psi(p) \in \cM$. The boundary
$\scri$ of $\widetilde{\cM}$ in $\cM$ represents `infinity' of the
physical manifold. The asymptotic region $\scri$ is in a sense unique
for all asymptotically flat space-times and defines a unique
`background' structure, on which one can define the mass of a
space-time and the gravitational radiation measured by an observer at
infinity.

Since one has immediate access to the geometric structure at $\scri$
in the unphysical manifold $\cM$, it is reasonable to solve the
conformally transformed Einstein equations directly in $\cM$.  Now the
problem arises, that the transformed Einstein equations are formally
singular at $\scri$. This problem was solved by Friedrich
\cite{friedrich81:_asymp_civp}, who proposed the conformal field equations, a system
of regular equations for the geometric structure on $\cM$. If there
exists a triple $(\cM, g, \Omega)$ satisfying the conformal field
equations then the pair $(\widetilde\cM, \tilde g)$, where
$\widetilde\cM = \{\Omega > 0\}$ and $\tilde g = \Omega^{-2}\,g$ is a
vacuum space-time.

Let us consider the abstract manifold $\cM$ with the additional
structure of a metric $g_{ab}$ and a metric compatible connection
$\nabla_{a}$. $T^{a}{}_{bc}$ and $R^{a}{}_{bcd}$ are the torsion and
curvature tensors, respectively, associated with $\nabla_{a}$.
Additional variables are the conformal factor $\Omega$, a 1-form
$\Sigma_{a}$, a scalar function $S$, a symmetric traceless tensor
field $\Phi_{ab}$ and a completely traceless tensor field $K_{abcd}$
with the symmetries of the Weyl tensor.

Then the conformal field equations read:
\begin{align}
 \label{15} T^{a}{}_{bc}&=0, \\
 \label{16} R_{abcd} - \Omega K_{abcd} + 2g_{c[a}\Phi_{b]d} -
 2g_{d[a}\Phi_{b]c}&\nonumber \\
- 4 g_{c[a}g_{b]d}\Lambda &=0, \\
 \label{17} 2\nabla_{[c}\Phi_{a]b} + 2g_{b[c}\nabla_{a]}\Lambda -
 K_{cab}{}^{d}\Sigma_{d} &= 0, \\
 \label{18} \nabla_{a}K^{a}{}_{bcd} &= 0,\\
 \label{19} \nabla_{a}\Omega - \Sigma_{a} &= 0,\\
 \label{20} \nabla_{a}\Sigma_{b} - g_{ab}S + \Omega \Phi_{ab} &=  0,\\
 \label{21} \nabla_{a}S + \Phi_{ab}\Sigma^{b} -
 \Omega\nabla_{a}\Lambda - 2 \Lambda\Sigma_{a}  &=  0,\\ 
 \label{22} 2 \Omega S - 2\Omega^{2}\Lambda - \Sigma_{a}\Sigma^{a}  &= 0.
\end{align}
The first equation states, that the torsion tensor $T^{a}{}_{bc}$
vanishes, so that $\nabla_{a}$ is the Levi-Civita connection of the
metric $g_{ab}$ and~(\ref{16}) is the decomposition of the curvature
tensor into its irreducible parts. Since one can show that the Weyl
tensor $C^a{}_{bcd}$ necessarily vanishes on $\scri$ it is written as
$C_{abcd}=\Omega K_{abcd}$. $\Phi_{ab}$ is the traceless part of the
Ricci tensor and $\Lambda$ is proportional to the Ricci scalar, $R =
24 \Lambda$.  Eq.~(\ref{19}) serves to get first order equations and,
therefore, must be regarded as the definition of $\Sigma_{a}$.
Eqs.~(\ref{20}) and~(\ref{22}) are the expressions for the conformally
transformed traceless part of the physical Ricci tensor and the
physical scalar curvature, respectively. Eqs.~(\ref{17})
and~(\ref{18}) are the conformally transformed Bianchi identities of
the physical manifold, which form a central part of the conformal
field equations.

Now we have to set up an initial value problem for the conformal field
equations. We do this by the usual $3+1$-decomposition. Since we are
working in the tetrad formalism, we pick a tetrad $e^{a}_{\mu}$ with
$g_{ab}=\eta_{\mu \nu} e^{\mu}_{a} \otimes e^{\nu}_{b}$, ($e^\nu_b$
being the dual frame of covectors) where $t^{a}=e^{a}_{0}$ is
orthogonal to the hypersurface $\mathcal{S}$ of constant coordinate
time $t$, so the manifold is sliced in the way $\cM = \mathbb{R}
\times \mathcal{S}$.  In a tedious but straightforward procedure we
get a system of symmetric hyperbolic evolution equations, together
with a system of constraints.

In this process, we also have to fix the gauge freedom. This is done
here by adding appropriate equations to the system, which contain 
four free functions, which determine the coordinates, six free functions 
corresponding to the freedom in the choice of the
tetrad. Since the conformal field equations are invariant under conformal
rescalings $\Omega
\rightarrow \omega\Omega, g_{ab} \rightarrow \omega^{2}g_{ab}$, where $\omega$
has to be strictly positive, there is
one additional function to fix the conformal factor.
The details of this hyperbolic reduction together with the fixing of the gauges are
described in \cite{friedrich96:_hyper_reduct} and \cite{frauendiener00:_confor_infin}.

Since our version of the conformal field equations is not suited to
treat space-like infinity $i^0$ in an appropriate way
(see~\cite{friedrich:_confor_einst_evolut} for a discussion of other formulations of the
conformal field equations) we choose as the initial data hypersurface
a hyperboloidal slice of $\cM$. This hyperboloidal slice intersects
$\scri^{+}$ in a space-like 2-surface, in contrast to an asymptotically
Euclidean hypersurface which reaches spatial infinity $i^{0}$. Since
a hyperboloidal slice is not a Cauchy surface for the entire space-time,
we are not able to calculate the entire space-time, but we can do semi-global
evolutions into the future. This is enough for our purposes, because we are mainly
interested in the asymptotic regions near and including $\scri$, where
we can read off the gravitational radiation emitted within the
manifold.

\subsection{Axisymmetry}

We assume axisymmetry of our physical space-time $\widetilde{\cM}$,
see \cite{carot00:_some} for a definition.  The assumption of axisymmetry is
motivated by the fact, that we can reduce the numerical problem from a
3D to a 2D problem and that this symmetry still allows gravitational
radiation, see \cite{ashtekarxanthopoulos78:_isomet} for the
symmetries which admit radiation for light-like asymptotically flat
space-times.  This allows us to simulate interesting radiating systems
with high accuracy already on small computers.  The symmetry will be
transferred to the conformal space-time $\cM$, if the conformal factor
$\Omega$ is also invariant under the symmetry
$\mathcal{L}_{\xi}\Omega=0$, which we assume explicitly by partly
using up the gauge freedom in $\Omega$.

The reduction from a 3D to a 2D problem is achieved by evolving only
within one of the hypersurfaces of constant $\phi$. All these
hypersurfaces are isometric and so it is enough to calculate all
fields on one of them.  The tricky part of axisymmetric systems is the
analytic and numerical treatment of the axis. From the analytical
side, one would like to exclude pathological things, like a massive
string on the axis. This means we would like to have a space-time,
which is regular on the axis. The regularity of the axis is usually
ensured by the condition of elementary flatness \cite{carot00:_some} on the
axis.  But this condition is very hard to implement
numerically. Fortunately, there is an equivalent condition of Wilson
and Clarke in \cite{wilsonclarke96:_elemen}, which simply states, that if the Killing
vector $\xi^{a}$ has the same form as in Cartesian coordinates and the
metric is differentiable in a neighborhood of the axis, which we have
to assume anyway for the numerics, then the axis is regular. This is
the reason, why we use Cartesian coordinates in our code. We evolve in
the hypersurface, where $\phi=0$, because there the Cartesian
$(x,z)$-coordinates coincide with the $(r,z)$-coordinates of polar
coordinates. In the following, we will call this hypersurface the
`evolution hypersurface'.

Apart from the axisymmetry we have an additional discrete symmetry
because the Killing vector $\xi^{a}$ is orthogonal to hypersurfaces of
constant $\phi$, so the reflection $i$ across the evolution
hypersurface given by $\phi \mapsto -\phi$ and hence $\xi^{a} \mapsto
-\xi^{a}$ is an isometry. Thus, by adapting the tetrad $e^{a}_{\mu}$
to the Killing vector, i.e., $e^{a}_{3} \sim \xi^{a}$, in the
evolution hypersurface we can achieve a reduction of the independent
tensor components. For this reduction we have to distinguish between
tensors fields, which are invariant under $i$ or which are mapped onto
their negative. In the first (second) class all tensor components
obtained by contracting with an odd (even) number of $e^a_3$ will be
zero. This reduces the problem from 65 free components to 35
components.

\section{Numerics}
\label{sec:numerics}

In this section we want to describe the numerical scheme, which we
used to solve the conformal field equations. Besides the conformal
field equations we tested the same scheme also on the scalar wave
equation in the Minkowski space-time. The numerical grid is uniform
and rectangular, where the outermost points are used for the boundary.
Since we use a non-staggered grid, $x=0$, is one of the boundaries of
our grid. All of these points lie on the axis of the space-time, which
is not a boundary of the space-time. Therefore, we have to treat these
points in a different way, which will be described in the subsection
on axisymmetry and the cartoon-method.

\subsection{Evolution}

For the evolution the Method of Lines is used. This is a semi-discrete
method for solving partial differential equations. The PDEs are
discretized in space to obtain a system of ODEs which are then
integrated in time along the lines in space-time given by fixing the
grid points. We use two standard methods for solving the ensuing ODEs:
a Runge-Kutta method and the multilevel Adams-Bashforth scheme, both
of fourth order in time. In addition, we have used the iterative
Crank-Nicholson (ICN)
scheme~\cite{teukolsky99:_crank_nichol_method_numer_relat}, which is
an explicit form of the implicit Crank-Nicholson scheme. These three
methods are all explicit, which makes them easy to implement in the
code. Nevertheless implicit methods would have better stability
properties, but compared to explicit methods they would be much
slower, because one has to solve a big system of nonlinear, coupled
equations at each time step.

The spatial discretization can be chosen between second order and
fourth order central stencils, but we use fourth order discretization
in all test runs, as there was no difference in the stability
properties of the method, only in the accuracy of the solution obtained.

\subsection{Axisymmetry and the Cartoon-Method}

We stated in the last chapter, that we use Cartesian coordinates, so
that the axis is regular.  Despite the fact that we are using non
adapted coordinates, we can reduce the problem from 3D to 2D by
evolving in the hypersurface, where $\phi=0$ and the Cartesian
coordinates $(x,z)$ coincide with the cylindrical coordinates
$(r,z)$. The problem, which arises now, is, that we will need
derivatives across this evolution hypersurface. We solve this by
applying an idea of Alcubierre et al. \cite{alcubierrebrandt99:_symmet}, which they call
the cartoon-method.  They have proposed, that instead of using the
symmetry in adapted coordinates $\pa_{\phi}T=0$, one can use the
symmetry as a tensor equation $\Phi^{*}T=T$ in Cartesian coordinates,
where $T$ is an arbitrary axisymmetric tensor. Because of the use of
Cartesian coordinates, we get in this way a regular axis and at the
same time we avoid the coordinate singularity of adapted coordinates
and the resulting singular components at the axis. One could certainly
cure the singularities of adapted coordinates by a regularizing
procedure for the components and the equations at the axis. But it
should be clear, that this change of the equations is a possible
source of instabilities and for our large system of 35 independent
components, it would have been at least tedious to check the behavior
of each component at the axis (see \cite{evans85}, \cite{thornburg96:_numer} for such
regularization procedures and~\cite{garfinkle:_simul} for a more recent
approach). As Cartesian coordinates are often used in 3D codes, the
cartoon method has the additional advantage that we can easily compare
the results of our code with that of 3D codes. In the following we
give the formulation of the cartoon-method in the tetrad formalism.

Since we are using a tetrad basis we not only have to ensure the
regularity of our coordinates, but also the regularity of our tetrad
at the axis. Additionally, we have the constraint, that one basis
vector $e^{a}_{i}$ has to be orthogonal to the evolution hypersurface,
so that we can use the discrete symmetry $i$ to reduce the number of
independent components.  Apart from these two conditions we can
arbitrarily fix the tetrad in our space-time. We first introduce an
invariant tetrad $f^{a}_{\mu}$, where $f^{a}_{3} \sim \xi^{a}$ and
$\Phi_{*}f^{b}_{\mu}=f^{b}_{\mu}$. Similar to adapted
coordinates, this adapted tetrad is singular on the axis. But we have
ensured, that our manifold is elementary flat at the axis, so we only
have to rotate the tetrad in the neighborhood of the axis, to a tetrad
which has a smooth limit while approaching the axis. So we make the
ansatz, that the tetrad is given by the rotation
$e^{b}_{\mu}=R_{\mu}{}^{\nu}f^{b}_{\nu}$, where
\begin{equation*}
    R_{\mu}{}^{\nu}=\left(\begin{array}{cccc} 1 & 0 & 0 & 0 \\
 0 & 1 & 0 & 0 \\
 0 & 0 &
    \frac{x}{\rho} & -\frac{y}{\rho} \\
 0 & 0 & \frac{y}{\rho} & \frac{x}{\rho}
    \end{array}\right).
\end{equation*}
Now we calculate the frame $f^{a}_{\mu}=f_{\mu}{}^{\nu}\pa^{a}_{\nu}$
with respect to the frame on the evolution hypersurface, where
$\pa^{a}_{\nu}$ are the coordinate vectors:
\begin{eqnarray*}
  (\Phi_{*} f^{a}_{\mu})(\Phi(p)) &=& f_{\mu}^{\nu}(p) (\Phi_{*}\pa^{a}_{\nu})(\Phi(p))
  = f_{\mu}^{\nu}(p) \Phi_{\nu}{}^{\kappa} \pa_{\kappa}^{a}(\Phi(p)), \\
  (\Phi_{*} f^{a}_{\mu})(\Phi(p)) &=& f_{\mu}^{\kappa}(\Phi(p))
  \pa_{\kappa}^{a}(\Phi(p)).
\end{eqnarray*}
We get the following relation between the invariant tetrad
$f^{a}_{\mu}$ at a point $p$ on the evolution hypersurface and a point
$\Phi(p)$ on the orbit of the isometry through $p$:
\begin{equation}
 f_{\mu}{}^{\kappa}(\Phi(p))=f_{\mu}{}^{\nu}(p)
  \Phi_{\nu}{}^{\kappa}.
\end{equation}
And together with the ansatz of our tetrad
$e^{b}_{\mu}=R_{\mu}{}^{\nu}f^{b}_{\nu}$, we get the tetrad components
anywhere in the manifold with respect to the components on the
evolution hypersurface:
\begin{eqnarray}
  c_{\nu}{}^{\kappa}(\Phi(p)) = R_{\nu}{}^{\mu}(\Phi(p))
  c_{\mu}{}^{\rho}(p) \Phi_{\rho}{}^{\kappa}.
\end{eqnarray}
Now for regularity of the tetrad, we only have to ensure that there
exists a uniform limit for the $c_{\nu}{}^{\kappa}$, when $p$
approaches the axis.  The coordinate vectors themselves have a regular
limit because of the usage of Cartesian coordinates. The necessary
behavior of the components $c_{\nu}{}^{\kappa}$ at the axis is then
built implicitly into the code.

The same procedure applied to an arbitrary tensor
$T^{ab\ldots}{}_{cd\ldots}$ with respect to the tetrad basis results
in the following relation for the components:
\begin{eqnarray}
 (\Phi^{*}T)^{\nu_{1} \nu_{2} \ldots}(\Phi(p)) =(R^{-1})_{\mu_{1}}{}^{\nu_{1}}(\Phi(p))
  \ldots T^{\mu_{1} \mu_{2} \ldots}(p).
\end{eqnarray}
All occurring geometric objects $T$ are axisymmetric $\Phi^{*}T=T$, so
we get the following relation:
\begin{eqnarray}
  T^{\nu_{1} \nu_{2} \ldots} (x,y,z) =(R^{-1})_{\mu_{1}}{}^{\nu_{1}}(x,y,z)
  \ldots T^{\mu_{1} \mu_{2} \ldots}(\sqrt{x^{2}+y^{2}},0,z).
\end{eqnarray}
In order to ensure a smooth limit for the tensors, this formula leads
to conditions for the components on the axis, which
are also built implicitly into the code.

This formula is also used to calculate the derivative $\pa_{y}$, which
points away from the grid.  The problem here is, that for the
derivative at grid point $(x_{i},z_{j})$ of the evolution hypersurface
in $y$-direction, we need the tensor components at points
$(\sqrt{x_{i}^2+ m^2 (\delta y)^2},0,z_{j})$, where $m$ depends on the
employed discretization. These points will in general not coincide
with a grid point.  So one has to use interpolation to calculate the
components there. In the paper on the cartoon-method by Alcubierre et
al. \cite{alcubierrebrandt99:_symmet} they propose to use Lagrange
interpolation. However, in \cite{frauendiener02:_some} we show for the scalar wave
equation, that this interpolation method results in an unstable
evolution scheme and propose alternative schemes, which are
stable. Unfortunately even with these alternative schemes we encounter
instabilities in the case of the conformal field equations.

\subsection{Boundary}

In our current approach, we set the boundary of our numerical grid
into the unphysical part of the conformal manifold.  As the physical
boundary $\scri$ is a characteristic hypersurface, analytically no
influence from the unphysical part can enter the physical part. Our
results show, that this is also true in our numerical treatment. So we
do not have to provide physically correct boundary conditions, like
e.g. constraint preserving boundary conditions, as one should if the
boundary would be in the physical part. Instead we only have to impose
numerical boundary conditions, which lead to a stable and convergent
numerical scheme.

This is not a much simpler task. We follow a standard
scheme~\cite{gustafssonkreiss72:_stabil} to get numerical boundary
conditions for a symmetric hyperbolic system, which imposes boundary
conditions on the ingoing local characteristic fields. The outgoing
local characteristic fields are determined by the data on the grid and
are extrapolated to the boundary point.  We only provide second order
boundary conditions, which are simpler and less computationally
expensive, than fourth order conditions. This has the obvious
disadvantage, that we can use only a second order spatial
discretization at the last inner grid point, which results in a global
third order scheme, as the numerical results indicate.

In order to get the local characteristic fields, one considers only
the principal part of the evolution equations $t^{a}\pa_{a} u = A^{i}
e^{a}_{i} \pa_{a} u$.  Then one determines the outward pointing unit
normal vector $n^{a}$ to the boundary and decomposes the principal
part into the parallel and orthogonal part to the normal vector:
$\pa_{t} u = (A^{i} n_{i} - t^{a}n_{a} E)n^{a} \pa_{a} u +
\textrm{tangential part}$, where $E$ is the identity matrix.  The
tangential part can be discarded, because it is only important, what
moves across the boundary. We determine the eigenvalues $\lambda$ and
eigenvectors $\Pi$ of the matrix $A^{i}n_{i} - t^{a}n_{a}E$ and
classify the local characteristic fields $\Pi$, by the sign of the
corresponding eigenvalue (this is the reason, why we have to ensure,
that the normal vector is outward pointing).
\begin{eqnarray*}
  \lambda > 0 &:& \textrm{ingoing},\\
  \lambda < 0 &:& \textrm{outgoing},\\
  \lambda = 0 &:& \textrm{tangential to the boundary}.
\end{eqnarray*}  
Now we impose boundary conditions on the ingoing local characteristic
fields of the following form:
\begin{eqnarray*}
 \Pi^{i}_{in} = \textrm{R}^{i}{}_{j} \; \Pi^{j}_{out} + g^{i}(t),
\end{eqnarray*}
where the matrix $R^{i}{}_{j}$ has to fulfill $\|RR^{T}\| \leq 1$ and
the function $g^{i}(t)$ can be specified with certain restrictions.
The values of the outgoing characteristic fields are extrapolated from
the data on the grid. The derivation of the boundary conditions seems
to indicate, that we have to extrapolate in the direction of the
normal vector. This is in general not well defined in a manifold and
as the normal vector has in general also time components, it may even
require to extrapolate in time not only in space, which is
obviously impossible. Therefore we take the most simple approach and
extrapolate along the grid lines.

The matrix $R^{i}{}_{j}$ acts like a reflection coefficient. As the
boundary has no physical meaning and is simply the boundary of the
numerical grid, reflection of the outgoing characteristic fields would
be wrong and we use highly absorbing boundary conditions, by setting
$R^{i}{}_{j}=0$.

At the moment we are not able to generate general initial data. For
this reason and because we want to test our numerical scheme we use
initial data generated from exact solutions. Therefore, we set
$g^{i}(t)$ to the value computed from the exact solution, when it is
not stated otherwise.

So far this has been the standard procedure for providing boundary
conditions, if a fixed background metric is available.  In our case,
the metric is determined by the tetrad components, for which we also
have to provide boundary values.  This means in principle we cannot
determine the normal vector at the boundary, because we do not know the
metric at the boundary. We solve this problem by extrapolating the
necessary tetrad components. From the extrapolated tetrad components
we calculate the eigenvalues of the characteristic fields. This should
not result in a significant error, because only the sign of the
eigenvalue is important. A bit more tricky is the transformation
from the characteristic fields to the physical fields. This yields
nonlinear coupled equations for the tetrad components, which occur in
the normal vector. We solve the occurring equations with the
Newton method.

The scheme, we have presented for providing boundary conditions, is a
generalization of the standard scheme of
\cite{gustafssonkreiss95:_time}, where one allows varying coefficients
in the symmetric hyperbolic principal part and in addition works
without a background metric. Therefore it is not clear, whether the
theorems derived for the standard case are still valid here.

\section{Results}
\label{sec:results}

In this section we describe the results, which we have obtained so
far. At first we describe the results for the scalar wave equation,
before we examine the conformal field equations. In both cases we use
the same numerical scheme. This means only the equations are
changed, the evolution and the treatment of the boundary and the
axis are the same.

In all cases the grid size $s$ is: $s=N_{x}(2 N_{x} -1)$, where
$N_{x}$ is the number of grid points in $x$-direction, and
$N_{z}=2N_{x}-1$ the corresponding number of points in the
$z$-direction. This ensures, that the origin lies on a grid point.

Errors $\Theta$ are measured with the $L_{1}$-Norm:
\begin{eqnarray*} 
\Theta(t) =
\frac{1}{s}\sum_{n=1}^{N_{x}-2}\sum_{m=1}^{N_{z}-2} ||\hat{u}(x(n),z(m),t)
- u_{n,m}(t)||_{1},  
\end{eqnarray*}
where $\hat{u}$ is the exact value of $u$ at $(x(n),z(m))$
and $u_{n,m}$ the value at the grid point $(n,m)$. In these
sums the boundary points are excluded. Here $u$ stands for either
the vector of all variables or the vector of all constraint 
components.

\subsection{Axisymmetric scalar wave equation}

At first we would like to discuss shortly the results of our code for
the axisymmetric scalar wave equation in flat space-time. As we have
written the scalar wave equation also in symmetric hyperbolic form, we
can use it as a testbed for our numerical scheme, especially for the
treatment of the boundary and the axisymmetry.

For the scalar wave equation Lagrange interpolation leads to an
unstable numerical scheme.  In \cite{frauendiener02:_some} we introduce a second
order interpolation procedure for the cartoon method, which results in
a stable second order evolution scheme. We have also found a fourth
order interpolation scheme, which seems to be stable, but we cannot
prove it yet. The resulting evolution scheme is third order accurate.
This is a result of our second order boundary treatment, which
changes our fourth order scheme globally to a third order scheme. 

The code was also tested on robust stability, which was proposed by
Winicour. This tests the scheme by evolving random initial data with
random boundary data whereby all possible solutions of the numerical
scheme will be excited. In our numerical tests we see no emerging
deviation from the mean after one million timesteps, where we stopped
the evolution.

These results show, that our numerical scheme is a suitable method for
solving axisymmetric systems as long as one can find an interpolation
method, which is compatible with the evolution scheme.

\subsection{Axisymmetric Conformal Field Equations}

Unfortunately the very good results from the scalar wave equation do
not transfer to the conformal field equations.  It was not possible to
get a stable evolution for any of the space-times, we have tested. But
this does not mean, that our results are physically meaningless. For
each space-time, where a regular point $i^{+}$ exists, it was possible
to evolve until $i^{+}$ and sometimes much further on.

As we have no general initial data yet, we tested our code with exact
solutions. The exact expressions for the fields were calculated by
GRTensorII \cite{grtensor} and then translated directly into C code.

\subsubsection{Minkowski Space-Time}

At first it may seem trivial to test the code with the Minkowski
space-time. But the conformal compactification of Minkowski space-time
has curvature and can even be time dependent, if one uses a time
dependent conformal gauge. We have tested three different
compactifications of the Minkowski space-time.

The first one is the standard compactification, as it is introduced in
textbooks:
\begin{eqnarray*}
  \label{first}
  \Omega &=& \cos(t) + \cos(r),\\ 
  g_{ab} &=& dt^{2} - dr^{2} - \sin^{2}(r) d\omega^{2},
\end{eqnarray*}
where $d\omega^{2}$ is the line element of the unit sphere. The second
one is a rescaling of the standard compactification, which has the property,
that $i^{0}$ is not a finite point:
\begin{eqnarray*}
  \label{second}
  \Omega&=&\cos(t)(1+r^{2}) + 1 - r^{2},\\
  g_{ab}&=&\frac{1}{4}(1 + r^{2}) dt^{2} - dx^{2} - dy^{2} - dz^{2},
\end{eqnarray*}
and the last one is the static compactification
\begin{eqnarray*}
   \label{third}
   \Omega&=&\frac{1}{2}(1-r^{2}),\\
   g_{ab}&=& \frac{1}{4}(r^{2}-1)^{2} dt^{2} + (x dx + y dy + z dz) dt\\
          && - dx^2 - dy^2 -dz^2. 
\end{eqnarray*}

\paragraph{Convergence}
We show convergence of our numerical scheme for the standard
compactification.
\begin{figure}[htbp]
 \begin{center}
   \leavevmode
   \epsffile{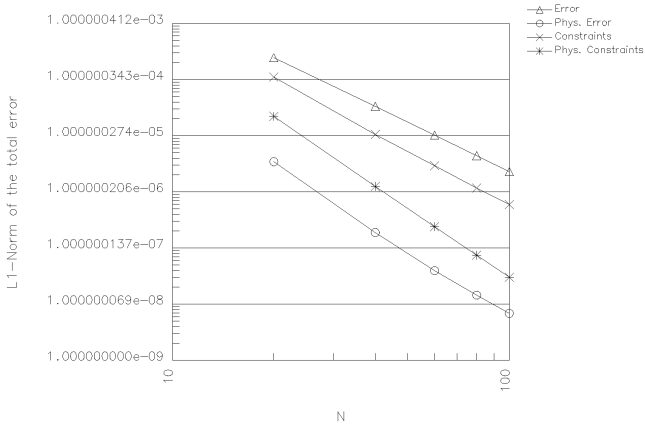}
  \end{center}
\caption{\label{fig:conv} Total error with respect to the exact solution
  for the standard compactification, for $N$=20,40,60,80 and 100
  points in the $x$-direction}
\end{figure}
In Fig.~\ref{fig:conv} we plot the error summed over all fields and
the constraint violation for the whole grid and for the physical part,
$\Omega \geq 0$, at coordinate time $t=2.1$ for increasing number of
grid points $N$, the initial time is $t_{ini}=1.8$. This yields a
convergence order for the total error of $2.903 \pm 0.002$, which is
in good agreement with our expectation of a third order scheme.  The
total constraint violation converges to zero with order $3.32 \pm
0.08$. The physical errors are both much smaller than the total error,
which comes from the fact, that the error is concentrated at the
boundary, which does not lie in the physical space-time.

A problem we face at the moment is, that the behaviour of the error
does not generalize to higher grid dimensions. There fast growing
instabilities arise at the boundary, which are not present at lower
resolutions. These instabilities dominate the total error after a short
time and lead to a crash at the boundary.

\begin{figure}[htbp]
 \begin{center}
   \leavevmode \epsffile{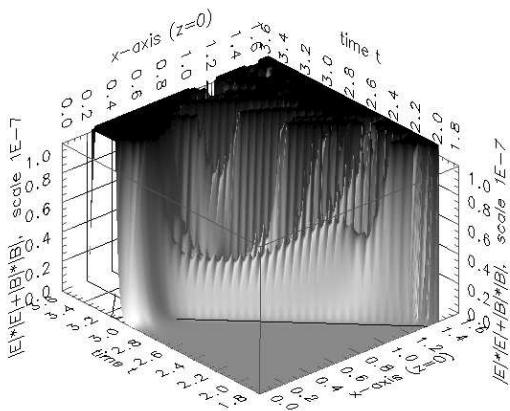}
 \end{center}
\caption{\label{fig:weyl} $C=|E_{ab}E^{ab} + B_{ab}B^{ab}|$ for the
  standard compactification with noise entering through 
  the boundary. The black line indicates $\scri$ ($\Omega=0$).}
\end{figure}

\paragraph{Numerical Properties of $\scri$}
The great advantage of the conformal approach, that the boundary of
the grid does not lie in the physical space-time, in opposite to the
other approaches in numerical relativity, can be further illustrated
with the following example.  We start with initial data from the
standard compactification and add during the evolution uniform random
noise to the exact values for the ingoing characteristic fields at the
boundary. The maximum of the noise is $\delta=0.005$, we use $40
\times 79$ grid points, initial time is $t_{ini}=1.8$. We plot in
Fig.~\ref{fig:weyl} the quantity $C=|E_{ab}E^{ab} + B_{ab}B^{ab}|$
composed out of the electric and magnetic part of the Weyl tensor. In
an undisturbed evolution $C$ remains zero all the time, because
$E_{ab} = B_{ab}=0$ in the initial data and this is preserved through
the numerical evolution.
In Fig.~\ref{fig:weyl} one can see two crucial points. First $\scri$
is not affected at all by the noise from the boundary. This means,
that $\scri$ is stable under small perturbations entering from the
unphysical part.  Second the noise from the boundary, which is also
added to the Weyl tensor, enters the grid, but does not enter the
physical space-time. Note that the scale of the plot is $10^{-7}$ and
the plot is cut off at $1.2 \times 10^{-7}$. The plot would have been $1000$
times bigger to cover the whole range of $C$. So $\scri$ is indeed a
characteristic hypersurface not only analytically but also with great
accuracy numerically. Near $i^{+}$ at $t=\pi$ a small amount of noise
enters the physical space-time, but this seems to be
related to the very poor resolution of the physical space-time in this
area.

Further support for the property of $\scri$ can be gained by the
following Fig.~\ref{fig:stand_noise} of the error and constraint
violation for the whole grid and the physical part.
\begin{figure}[htbp]
 \begin{center}
   \leavevmode \epsffile{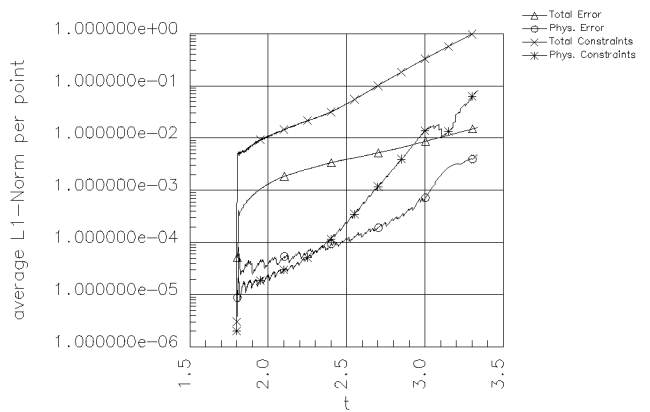}
 \end{center}
\caption{\label{fig:stand_noise} Error and constraint violation for
  the standard compactification with noise entering 
  through the boundary.}
\end{figure}
It can clearly be seen, that the physical error is much smaller than
the total error, the same is valid, but not with the same accuracy,
for the constraint violation. The reason for this might be, that one
calculates the physical constraint violation at points, where $\Omega
\geq 0$, but through the discretization for the derivatives one gets
an influence from points outside of the physical space-time. Both the
physical error and the constraint violation in the physical part grow
rapidly, while approaching $i^{+}$ at $t=\pi$, which has the reason in
the low number of physical grid points shortly before $i^{+}$. This
behavior of the error and the constraint violation is not limited to
this special case. Also for all other tested space-times, one can see,
that the error in the physical part is small relative to the error in
the unphysical part. We have made several movies, where one can see,
that the error stays outside of $\scri$. These movies are
available on our home page \cite{Home}.

\paragraph{Reproducing the static Einstein universe}

\begin{figure}[htbp]
\begin{center}
  \leavevmode \epsffile{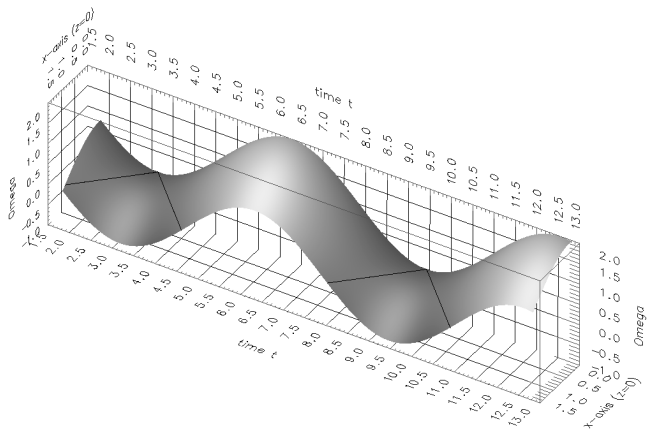}
 \end{center}
\caption{\label{fig:stand_omega} The conformal factor $\Omega$ for the
  standard compactification at $z=0$. The black line indicates $\scri$
  ($\Omega=0$)}
\end{figure}
In Fig.~\ref{fig:stand_omega} we plot the conformal factor of the
standard compactification, where we give the exact values for the
ingoing characteristic fields during the evolution. It can be seen,
that the physical space-time bounded by $\scri$ shrinks to a point
$i^{+}$, so it is no problem to reach $i^{+}$ and continue in a smooth
way. In principle it is now physically meaningless to continue the
evolution, because the whole physical part has been calculated.  But
from the numerical point of view it is interesting to see, for how
long it is possible to evolve further, before the code crashes.  The
plot shows, that it is possible to calculate additional one and a half
copies of Minkowski space-time.  As the Minkowski space-time is
embedded in the static Einstein universe, we obtain nothing else but a
part of the static Einstein universe.

\paragraph{Conformal Gauges}

As we have pointed out in \ref{sec:setting} in the conformal picture there
is one additional gauge freedom apart from the coordinate gauge and 
the tetrad gauge.  We are allowed to
rescale the conformal factor and the metric according to $\Omega
\rightarrow \omega\Omega, g_{ab} \rightarrow \omega^{2}g_{ab}$, where
$\omega$ has to be strictly positive, so that the asymptotic structure
is conserved by this conformal rescaling.

For the second conformal manifold \eqref{second} we tested different
conformal gauges. Some of these gauges lead to a very fast exponential
growth of the error and were therefore useless. The most successful
choice was the gauge:
\begin{eqnarray*}
 \omega = \frac{1}{1+x^2+z^2},
\end{eqnarray*}
which made many components spatially constant. It is therefore not
too surprising, that this solution had the smallest exponential growth
of the error. Motivated by this gauge, we tried another time dependent
one, which should model a nearly constant factor $\omega$ in the
physical space-time and then a decaying factor in the unphysical
space-time in order to have very small values at the boundary:
\begin{eqnarray*}
 \omega=\frac{1}{1+(x^2+z^2)t}
\end{eqnarray*}
This function has the serious drawback, that it forms a sharp peek at
the axis as time increases. This results in also very peeked
solutions, which leads to a more rapid exponential growth of the
error.

\begin{figure}[htbp]
\begin{center}
  \leavevmode \epsffile{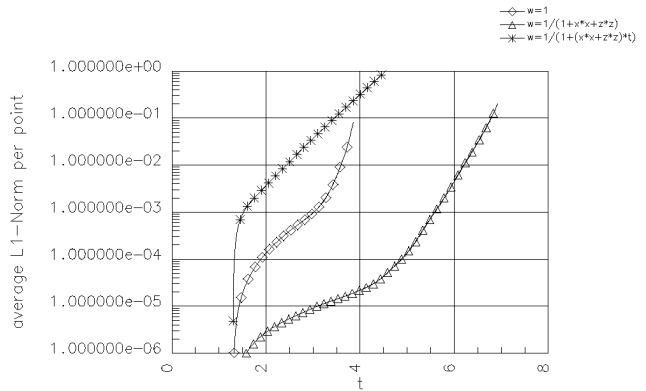}
 \end{center}
\caption{\label{fig:gauges} The effect of different conformal gauges
  on the total error for the second compactification 
  of Minkowski \eqref{second} space-time, grid: $40\times79$, $t_{ini}=1.3$}
\end{figure}

These results are illustrated in the next plot (Fig.~\ref{fig:gauges}).
As the exponential growth is different for the chosen gauges, it
seems, as if different exponentially growing solutions are excited. We
do not know, whether this difference emerges already on the level of
the conformal field equations or on the level of the numerical
implementation. As the exponential growth is a serious problem for
long term runs, this has to be further examined in the future.

\paragraph{Comparison of the three compactifications}

\begin{figure}[htbp]
\begin{center}
  \leavevmode \epsffile{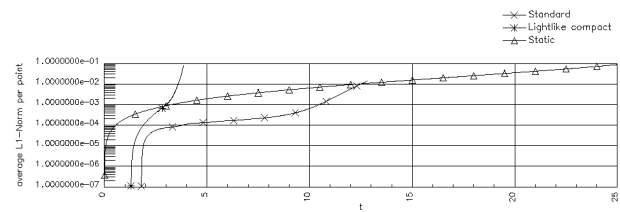}
 \end{center}
\caption{\label{fig:vergleich} The long term behavior of the total
  error of three different compactifications
  of Minkowski space-time, the marks are set every 1000 time steps,
  grid size: $40\times79$}
\end{figure}

Before we conclude the section on the Minkowski space-time, we would
like to illustrate the long term behavior of the three different
compactifications in Fig.~\ref{fig:vergleich}.
The absolute value is not comparable, because the step size is not the
same for the three cases. But the exponential growth can be compared,
because in principle this should not depend on the step size. As
expected, the fully static solution runs for the longest time. The big
difference in the stability between the standard compactification and
the light-like compact one seems to arise, because of the greater
spatial variance of the light-like compact solution.

\subsubsection{Boost-axisymmetric solution of Bi\v{c}\'ak and Schmidt}

As a further testbed for our code, we chose an asymptotically flat
solution from the general class of boost-axisymmetric space-times
analysed by Bi\v{c}\'ak and Schmidt \cite{bischmidt89:_boost_rotat}.  Since we are
working in the conformal picture it is necessary, that the solution
admits a regular $\scri^{+}$ and is regular on the axis. Indeed there
is a whole family of such solutions in the class of boost-axisymmetric
solutions, which fulfills these conditions, except for two singular
points on $\scri^{+}$, see \cite{bicak83:_einst}. This family of solutions
describes two self-accelerating particles and the singular points on
$\scri^{+}$ are the places, where the particles hit $\scri^{+}$.
\begin{eqnarray*}
 ds^{2} &=& e^{\lambda} \left( \frac{(x dx + y dy)^{2}}{x^2+y^2} - \frac{(t
   dt - z dz)^{2}}{t^2-z^2} \right) \\
          &&+ e^{-\mu}\frac{(x dy - y dx)^{2}}{x^2+y^2} +
          e^{\mu}\frac{(t dz - z dt)^{2}}{t^2-z^2} \\
 \Omega &=& - t^2 + x^2 + y^2 + z^2
\end{eqnarray*} 
with
\begin{eqnarray*}
 \mu(x^2+y^2,t^2-z^2)    &=&   m (t^2 - z^2 - x^2 - y^2),\\
 \lambda(x^2+y^2,t^2-z^2)&=&   m^2 (x^2+y^2)^2 - m (x^2 + y^2 + t^2 - z^2),
\end{eqnarray*}
where $m$ is a free parameter. In the limit $m \rightarrow 0$ the
solution approaches Minkowski space-time.  In order to avoid the
singularities in the evolution we chose our hyperboloidal initial data
hypersurface in the future of the singular points and so we are able
to calculate the rest of the space-time up to $i^{+}$.

\begin{figure}[htbp]
\begin{center}
  \leavevmode \epsffile{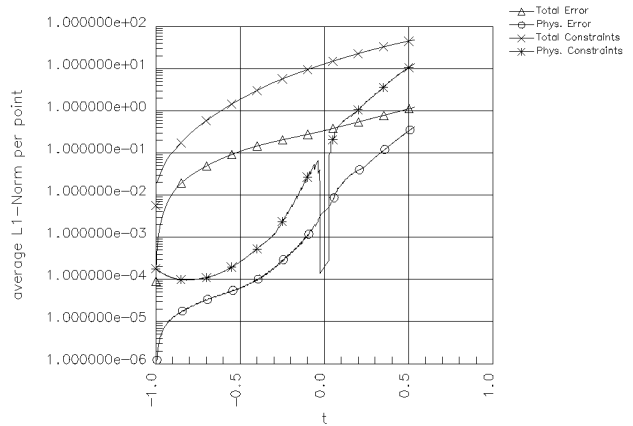}
 \end{center}
\caption{\label{fig:bic_error} Error and constraint violation for the
  boost-axisymmetric solution with strong data $m=1.75$, grid size:
  $40\times79$}
\end{figure}

We first show an error plot in Fig.~\ref{fig:bic_error} for a run with
$m=1.75$ and a grid of $40\times79$ points, which shows, that there is no
problem to reach $i^{+}$ at $t=0$ and to evolve further.
The high difference between total and physical error arises, because
of the very high values near the grid boundary.  This error does only
enter the physical space-time to a certain degree, which shows again
the advantage of the conformal approach to have the boundary in the
unphysical part of the space-time together with a characteristic
hypersurface $\scri$, which separates the unphysical from the physical
part.

\begin{figure}[htbp]
\begin{center}
  \leavevmode \epsffile{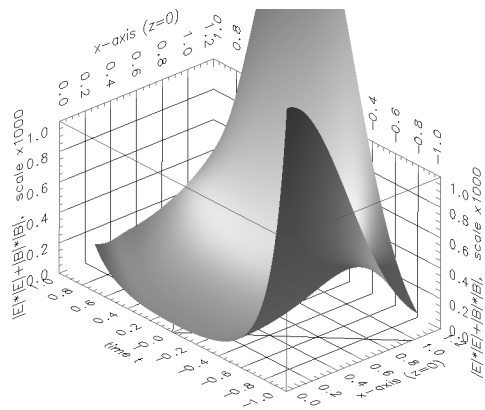}
 \end{center}
\caption{\label{fig:weyl_bicak} $C=|E_{ab}E^{ab}+B_{ab}B^{ab}|$ for
  the boost-axisymmetric solution with $m=1.75$. Note 
  the scale of 1000. The black line indicates $\scri$} 
\end{figure}

To show that this case is non-trivial in the sense that the curvature
does not vanish we plot in Fig.~\ref{fig:weyl_bicak} the quantity
$C=|E_{ab}E^{ab} + B_{ab}B^{ab}|$.
The scale of the plot indicates, that we are really dealing with data
in the non-linear regime and that the numerical scheme can deal with
large gradients in the data.

\section{Summary}
\label{sec:summary}

In this article we have presented a new code for simulating
axisymmetric, isolated systems in general relativity.  We have
implemented the axisymmetry by adapting the cartoon method to the
tetrad formalism and formulated boundary conditions, which are
compatible with the symmetric hyperbolic principal part of our
evolution equations.  We have tested the numerical scheme by applying
it to the axisymmetric scalar wave equation. The results show, that we
can solve the wave equation in a stable and convergent way, by
choosing a suitable interpolation scheme for the cartoon method. The
fact that the code is still unstable for the quasilinear conformal
field equations needs to be examined in further detail.

The numerical test for the conformal field equations with
hyperboloidal initial data from different compactifications of the
Minkowski space-time and strong data from a family of
boost-axisymmetric space-times show, that the semi-global simulation
of space-times until $i^{+}$ and further on is possible. In addition,
we have demonstrated, that $\scri$ is also numerically a
characteristic hypersurface and acts as a boundary between physical and
unphysical space-time with good accuracy.

There remain several tasks. We need to develop modules to extract the
radiation on $\scri$ and to determine the Bondi mass. These are
straightforward implementations of the algebraic transformation
between our tetrad and the tetrad adapted to $\scri$. Furthermore, we
need to improve our boundary conditions to be at least third order
accurate so that we do not loose one order of global accuracy as we do
now. An analysis of the initial boundary value problem needs to be
carried out (possibly along the lines of \cite{friedrichnagy98:_ibvp})
in order to be able to provide boundary conditions which are
compatible with the constraints. This would minimize the influence of
constraint violating modes generated at the boundary.

\begin{acknowledgments}
  We are grateful to Bernd G. Schmidt for his help in obtaining and
  preparing the exact expressions for the boost-rotation symmetric
  solution. 
\end{acknowledgments}


\end{document}